# Durable Agency Support for Exoplanet Catalogs and Archives

Principal Author: Joshua Pepper[1] (joshua.pepper@lehigh.edu)
Contributing Authors: David Ciardi[2], Todd Henry[3], Susan Mullally[4]

Thematic Areas: Activity, Exoplanets, Catalogs, Archives

Endorsing Authors
Elisabeth Adams[5], Daniel Angerhausen[6], David Ardila[7], G. Bruce Berriman[2], Douglas A. Caldwell[8], Tiago Campante[9], Joleen Carlberg[10], Jessie Christiansen[11], Karen Collins[12], Scott Fleming[10], Dawn Gelino[2], Yasuhiro Hasegawa[13], Brian Jackson[14], Wei-Chun Jao[15], Stephen Kane[16], Jonathan Labadie-Bartz[17], Patrick Lowrance[11], Michael Meyer[18], Gijs Mulders[19], Martin Paegert[20], Matthew Penny[21], Peter Plavchan[22], David Rodriguez[10], Sam Ragland[23], Emily Rauscher[18], Joseph E. Rodriguez[24], Dmitry Savransky[25], Jason Steffen[26], Rachel Street[27], Angelle Tanner[28], Stuart F. Taylor[29], Margaret Turnbull[8]

[1] Lehigh University, [2] Caltech/IPAC-NexScI, [3] RECONS Institute, [4] STScI,
[5] Planetary Science Institute, [6] Bern University, [7] Jet Propulsion Laboratory, [8] SETI Institute,
[9] IA (Portugal), [10] STScI, [11] Caltech/IPAC, [12] CfA/SAO, [13] JPL/Caltech, [14] Boise State University,
[15] Georgia State University, [16] UC Riverside, [17] University of Sao Paulo,
[18] University of Michigan, [19] University of Chicago, [20] Center for Astrophysics,
[21] Ohio State University, [22] George Mason University, [23] W.M. Keck Observatory,
[24] CfA | Harvard & Smithsonian, [25] Cornell University, [26] UNLV, [27] Las Cumbres Observatory,
[28] Mississippi State University, [29] Participation Worldscope

## Abstract

Many projects in current exoplanet science make use of catalogs of known exoplanets and their host stars. These may be used for demographic, population, and statistical studies, or for identifying targets for future observations. The ability to efficiently and accurately conduct exoplanet science depends on the completeness, accuracy, and access to these catalogs. In this white paper, we argue that long-term agency support and maintenance of exoplanet archives is of crucial importance to achieving the scientific goals of the community and the strategic goals of the funding agencies. As such, it is imperative that these facilities are appropriately supported and maintained by the national funding agencies.

## Context

Almost all current scientific studies of exoplanets depend on reliable catalogs of exoplanets and their respective system parameters. Demographic studies (i.e., analysis of the group properties of planetary systems to understand their origins and distributions), as well as selecting targets for intensive observations (e.g., atmospheric characterization) all rely on having well-curated and up-to-date properties of known exoplanets. The integrity of the exoplanet databases and catalogs in turn depends on the reliable archiving of exoplanet information. The curation of data and parameters relevant for the known exoplanetary host systems need to accurately reflect the content of the refereed literature or missions data products. Such archiving must be done carefully and diligently and relies heavily on experienced scientific and engineering staff, as well as on the underlying hardware and software infrastructure, necessary to manage the data and provide access to the data through public-facing interfaces.

The main exoplanet catalogs used by researchers are the NASA Exoplanet Archive (NEA), The Exoplanet Encyclopedia (exoplanet.eu) and The Exoplanet Orbit Database (exoplanets.org; Han et al. (2014)). Significant portions of major policy papers, including the Exoplanet Science Strategy (ESS)[1], draw upon these archives as a basis for evaluating the exoplanet field. This can be seen by examining the sources of information considered in Section 1 of the ESS, which reviews the "The State of the Field of Exoplanets". The following sections of that document draw information from existing archives:

- A number of key figures draw data directly from the NEA (Figs 2.1, 2.7, 3.1, 4.1), or are based on papers that drew data from the NEA (Fig 2.3) or the EOD (Fig 2.2).

[1] https://www.nap.edu/catalog/25187/exoplanet-science-strategy

- The frequency of hot Jupiters is taken from Wright et al (2012), which draws its data from exoplanets.org.
- The finding that giant planets are less common around M dwarfs cites Johnson et al. (2010), which bases its data on exoplanets.org, and Wang and Fischer (2015), which obtains stellar parameters from the NEA.
- Most of the planet frequency papers draw upon the Kepler Candidate or Confirmed Exoplanet tables at the NEA. Kepler project deliveries (Burke et al (2015) and Thompson et al (2018)) regarding the number and properties of planetary candidates are made available to the public through the NEA, while other population studies use the data on the NEA as the basis for analysis, including seminal papers such as Dressing & Charbonneau (2015) and Fulton et al (2017).

Data in these archives are the basis from which we draw conclusions about the underlying frequency and parameter distribution of planets and the associated host stars as well as the efficiency of various planet detection techniques. The archives are also crucial for selecting the best targets for characterization surveys which conduct detailed observations of individual planets, including for atmospheric characterization.

Right now the most widely used archive of exoplanet information used for studying exoplanet demographics and populations is the NASA Exoplanet Archive (NEA), which has been established and funded by NASA specifically to be this sort of resource for the scientific community and NASA missions. While there exist other lists of known exoplanets, such as the Exoplanet Encyclopedia and exoplanets.org, neither project maintains the level of completeness of the NEA. That statement is based on anecdotal experiences for a number of scientists working with data from the archives. It is also worth noting that exoplanets.org stopped incorporating new planets in June 2018.

The only detailed, quantitative comparison of the different exoplanet catalogs is found in Bashi, Helled, & Zucker (2018). That paper examined the statistical differences between catalogs, but did not attempt to investigate the reliability or the quality of the catalogs. That is, it did not attempt to determine whether the information in the catalogs was accurate or complete. If the community continues to maintain and use multiple catalogs in the coming years, such a comparison will be essential.

## Challenges

While individual missions have clearly specified and supported data archives, curated sets of derived and synthesized information are less substantially supported. Both IPAC and MAST maintain sets of high-level science products (HLSPs), in addition to the lower level mission data products, but those are often housed as specific, uniform science products supporting broader and more diverse science cases. The catalog of all known and suspected exoplanetary systems is a larger and more heterogeneous set of information, determined from multiple data types (imagery, photometry, spectroscopy, astrometry) using many different analysis techniques and spanning much of electromagnetic spectrum from X-rays to radio. Maintaining such a catalog, and the associated data, is a more difficult endeavor, and that difficulty should be reflected in the resources allocated to that work, commensurate with its importance.

Maintaining such an archive is not simply a matter of populating a database from a mission, housing a HLSP, or scraping numbers from the published literature. The information in the archive must be culled and assessed from the existing literature as individual planets are discovered, their parameters measured, their host stars characterized, and then as additional planets are discovered in the same system. The exoplanet archives need to be as fluid as the scientific process. The information is acquired in an unpredictable and heterogeneous way, paper by paper, with all the attendant ambiguity of new science results. The archive staff have to effectively adjudicate competing claims about the values of a given parameter, or even the existence of a particular planet. This process is much more than the transmission of numbers from the literature - it is the synthesis and curation of diverse information in a scientifically meaningful manner.

The challenge is even more difficult in terms of deciding what kinds of objects should be included in catalogs. Traditionally, exoplanets are considered to be compact objects with masses lower than that of brown dwarfs (roughly 15 $M_J$). However, it might be valuable for the community to include brown dwarfs in such catalog as well, especially for brown dwarfs that are bound to stars. It is not clear whether it makes sense to include unbound planet-mass objects in such catalogs, and what the standards should be for inclusion, especially if dynamic mass confirmation for such objects is not available. Beyond such questions stretches the broader debate over whether planetary status should be ascribed to present-day object mass or formation pathway, and if so what that means for catalog listings.

An additional difficulty, faced by all exoplanet catalogs, is the absence of community standards for reporting planet information - both in terms of what is reported as well as in what form items are reported. For example, some papers yield full orbital and system parameter solutions while others only provide a handful of parameters, and often there are multiple solutions for the same system within the same paper (created under different assumptions or analysis methods). Additionally, different papers often report results in different ways without clearly defining them (e.g., transit depths as magnitudes, millimagnitudes, flux ratios, or percentage flux ratios). Planet ephemeris times ($T_0$) are sometimes reported as the time of periastron passage, or as the time of mid-transit. Currently, the catalog curators must work to assess and identify the correct formats, and often correct mistakes published in the literature prior to ingestion into the catalogs.

Corresponding challenges exist for the curation and archiving of radial-velocity (RV) data, as described in a white paper by Steffen et al. (2018) for the NAS Exoplanet Science Strategy effort. Standards for data structures and types, preservation of raw spectra, proper recording of metadata, and associated assumed or extracted stellar parameters are all needed to allow future researchers to verify or expand upon the results from existing RV data sets. Precision radial velocity is one of the most important techniques for the discovery and characterization of exoplanetary systems (e.g., Ciardi et al. 2019) and long term curation and archiving of these data is needed (Steffen et al. 2018). As an example, the NEA currently serves more than 1000 published radial velocity curves for approximately 70% of the known radial velocity planets, but the effort to maintain this will grow with time as TESS finds more systems suitable for radial velocity follow-up and orbital solution and planetary mass determinations.

It should also be noted that in addition to the properties of the planets and their orbits, the careful amalgamation of data on the host stars is essential for understanding the planets. Knowledge of the stars' astrometric, photometric, and spectroscopic properties is essential for planning further observations of the systems, and their physical properties (mass, radius, luminosity, distance, space motion, metallicity, multiplicity) must be understood to understand the planets (see the program of the Know Thy Star, Know Thy Planet conference[2]).

The maintenance of an exoplanet archive is a fundamentally time-intensive process, requiring significant work by scientists, programmers, database architects, and user experience architects. NASA established the NEA, which is currently the most-used archive for demographic studies of exoplanets. As is true for other archives, work at the NEA is constrained by the available resources, and needs to be sufficient to support the full range of data products, user tools and infrastructure necessary for community use now and in the future.

There is also an issue of permanence and durability. Because the detection of exoplanets often takes place on timescales of multiples of the orbital periods (at least for transit, astrometric, and RV surveys), the observational data may well span decades. Discovery and characterization of planetary systems often relies on datasets, published and unpublished, that are years in the making and from various telescopic resources. The existence of these data and these data themselves must be preserved in an accessible format over such timescales, especially if it is to be used to select targets for observations with future resources in the 2020’s, 2030’s, or 2040’s. Steffen et al (2018) describe why proper archiving of survey data is so important, and those same arguments apply to other survey data, including for transit, microlensing, astrometric, and imaging surveys. Without proper archiving and curation of these data, in addition to the data and parameters of the discovered exoplanets, the demographics and properties of the exoplanetary systems, and our place in those distributions, can not be fully understood.

[2] http://nexsci.caltech.edu/conferences/2017/knowthystar/

Finally, the archive needs are expected to grow over time. As in nearly all new science areas, the number of new discoveries typically grows in a geometric or exponential manner. While exoplanet catalogs now contain more than 4,000 confirmed planets and another 1,000 candidates, current and upcoming missions will increase those numbers by one to two orders of magnitude, including expected detections from TESS (Barclay, Pepper, & Quintana, 2018), Gaia (Perryman, et al. 2014), WFIRST (Penny, et al. 2019), and PLATO[3]. As the search for earth-sized and earth-like planets and the desire to understand our own Solar System in the context of the galactic exoplanet population intensifies, the growth and rate of growth of the known exoplanets will also intensify (see Fig 1 below).

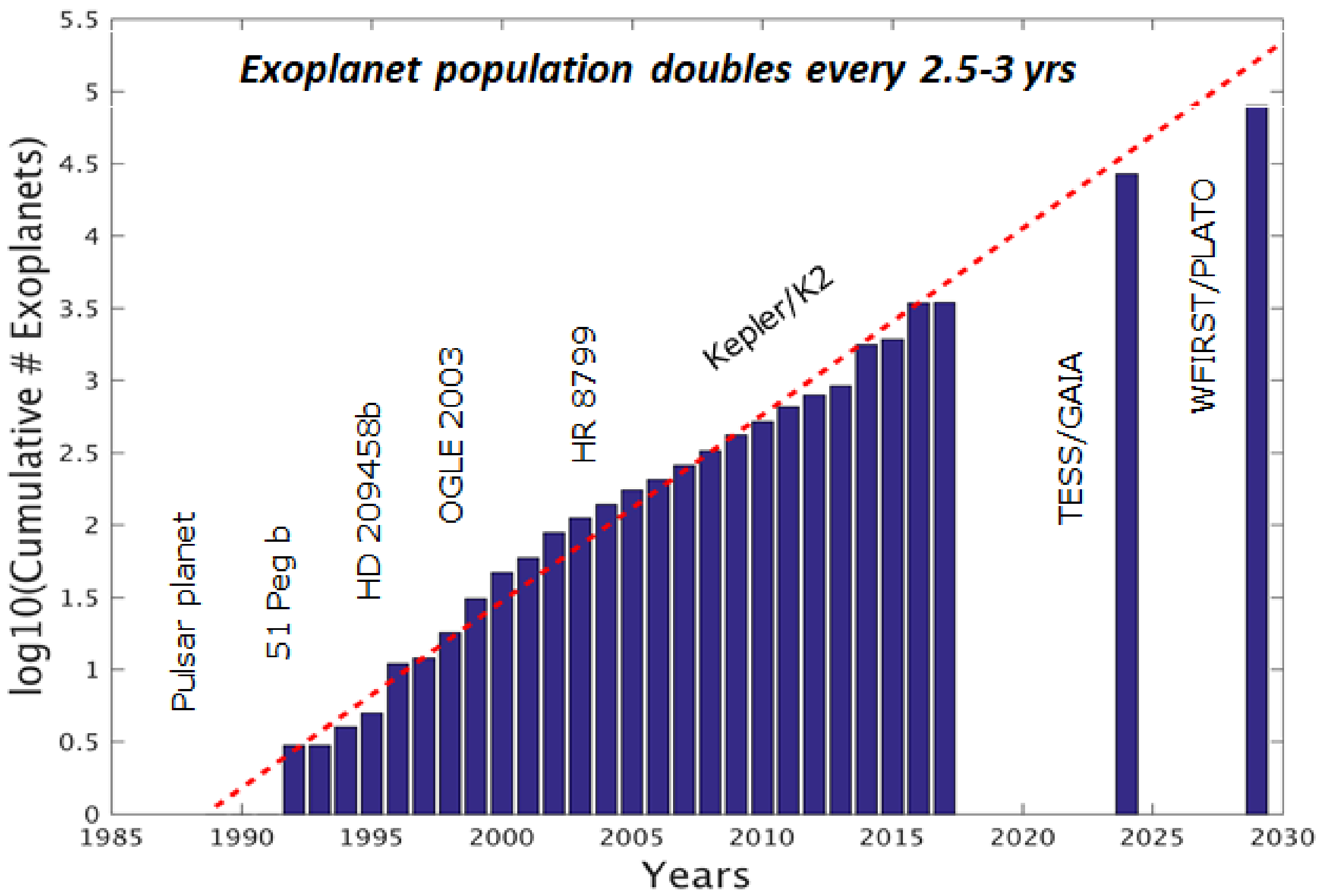

Figure 1: Growth in rate of exoplanet discovery through the present, and including predictions from upcoming missions. (Data Source: NASA Exoplanet Archive, figure from C. Beichman )

---

[3] https://platomission.com/2018/06/04/planet-yield/

## Recommendations

1. Funding agencies must provide support and resources for the exoplanet archives and curated catalogs commensurate with the needs of the scientific community. Resources include sufficient staff for data management, data ingestion, tool development, cross-connection to various agency archives, as well as user support and the necessary infrastructure to meet the demands of the services.
2. Working with the existing archive staffs, develop a set of community standards that can enable the scientific needs of the community and the technical needs of the archives.
    - Defining what is meant by “confirming” or “validating” exoplanet candidates for the purposes of catalog inclusion using data driven metrics
    - Recommending a range of parameters that a compact object should meet to be included in an exoplanet catalog.
    - Develop a unified terminology and parameter format for listing exoplanet system parameters in published venues, including journal articles
    - Define a common format for data files and inclusion of metadata for standard types of exoplanet measurement data, including photometry for transit surveys and spectroscopy for RV surveys
3. Funding agencies that support exoplanet discovery projects should require that such support be contingent on the completeness, rapidity, and quality with which they provide the project results to public archives.
    - A key part of this recommendation is that data management plans be evaluated as part of the core science and deliverables of a proposal. Proper data releases and the associated data management plans should be treated as an integral part of a scientific project, as much as conducting observations, computational modeling, or analyzing data, and not as an ancillary bookkeeping step.
    - Agencies should ensure that investigators can newly propose explicitly for funding to support the public archiving and release of data from previously-funded projects, if such activities were not funded in the original proposals. At the same time, proposals for funding new work should require data management plans that include the timely and complete release of all data products from the project.
    - Data management plans for large projects should be developed in contact with the relevant NASA-supported archive at the time of writing and receiving the grant.